\documentclass[12pt]{article}
\usepackage{amsfonts}
\usepackage{amssymb}

\usepackage[mathscr]{eucal}
\usepackage{amssymb,latexsym}
\usepackage{graphicx,verbatim}
\usepackage{epstopdf}
\usepackage{mathrsfs}

\usepackage[colorinlistoftodos]{todonotes}

\usepackage{amsmath}
\usepackage{amsfonts}
\usepackage{amssymb}
\usepackage{psfrag}
\usepackage{accents}

\newcommand\be{\begin{eqnarray}}
\newcommand\ee{\end{eqnarray}}
\newcommand\ben{\begin{enumerate}}
\newcommand\een{\end{enumerate}}
\newcommand\bit{\begin{itemize}}
\newcommand\eit{\end{itemize}}
\newcommand{\Lie}[0]{{\cal L}\, }

\def\lp{\ell_{\rm planck}}
\def\G{\mathbf{\Gamma}}
\def\O{\mathbf{\Omega}}
\def\X{{\bf X}}
\def\H{{\bf H}}

\def\scri{\mathscr{I^+}}
\def\B{\mathscr{B}}
\def\E{\mathscr{E}}
\def\W{\mathscr{W}}

\def\DH{\mathscr{H}}
\def\k{\kappa}
\def\f{\frac}
\def\a{a}
\def\IH{\Delta}
\def\l{\ell}
\def\t{\tilde}
\def\qo{\mathring{q}}
\def\Yo{\mathring{Y}}
\def\rmd{\rm d}

\def\ni{\noindent}

\makeatletter
 \newcommand{\pback}[1]{{
   \let\@rrow=\leftarrowfill
   \mathchoice{\AIN@stemPullBack{#1}{\@rrow}}{\AIN@stemPullBack{#1}{\@rrow}}
     {\AIN@indxPullBack{#1}{\@rrow}}{\AIN@indxPullBack{#1}{\@rrow}}}
   \vphantom{#1}}

 \newcommand{\AIN@stemPullBack}[2]{
   \vtop{\mathsurround=0pt
   \ialign{##\crcr$\textstyle{#1}\strut$\crcr
     \noalign{\kern-0.4ex\nointerlineskip}{\tiny#2}\crcr}}}

 \newcommand{\AIN@indxPullBack}[2]{
   \vtop{\mathsurround=0pt
   \ialign{##\crcr\hfil$\scriptstyle{#1}$\hfil\crcr
     \noalign{\kern+0.4ex\nointerlineskip}{\tiny#2}\crcr}}}

\makeatother

\def\S{\mathbb{S}}
\def\R{\mathbb{R}}
\def\M{\mathbb{M}}

\newcommand*{\scrip}{\ensuremath{\mathscr{I}^{+}}} 

\newcounter{mnotecount}[section]

\begin{document}

\title{BLACK HOLE HORIZONS\\ AND THEIR MECHANICS}

\author{Abhay Ashtekar \\
Institute for Gravitational Physics and Geometry \\ Physics
Department, Penn State, University Park, PA 16802-6300 }

\maketitle

\begin{abstract}

Black holes are often characterized by event horizons, following the literature that laid the mathematical foundations of the subject in the 1970s. However black hole event horizons have two fundamental conceptual limitations. First, they are defined only in space-times that admit a future conformal boundary. Second, they are teleological; their formation and growth is not determined by local physics but depends on what could happen in the distant future. Therefore, event horizons have not played much of a role in the recent theoretical advances that were sparked by discoveries of the LIGO-virgo collaborations. This article focuses on  quasi-local horizons that have been used instead. Laws governing them --mechanics of quasi-local horizons-- generalize those that were first found using event horizons.  These results, obtained over the last two decades or so, have provided much insight into dynamical predictions of general relativity in the fully nonlinear regime. The article summarizes the deep and multi-faceted interplay between geometry and physics that has emerged. Conceptually, quasi-local horizons also play a key role in the discussion of the quantum evaporation of black holes. However, due to space limitations, this application is only briefly discussed in Section 6.

\end{abstract}

\section{Introduction}
\label{s1}

By now, gravitational wave observations have established that black holes (BHs) with tens of solar masses are ubiquitous in our universe, and the Event Horizon Telescope (EHT) has provided us with images of light rings around two supermassive black holes. But what exactly is a BH? What exactly is it that forms as a result of a gravitational collapse and evaporates due to quantum radiation? The common answer is:\, \emph{Event Horizons}(EHs). Much of the rationale behind this conviction comes from the laws of BH mechanics that dictate the behavior of EHs in equilibrium, under small perturbations away from equilibrium, and in fully dynamical situations. While these laws are consequences of classical general relativity alone, they have close similarity with the laws of thermodynamics. The origin of this seemingly strange coincidence lies in quantum physics. Thanks to these insights, in the 1970s and 1980s, EHs were taken to be synonymous with BHs. 

However, EHs are rather enigmatic. In particular, they have \emph{teleological} properties. For example, an EH may be forming and growing in your very room \emph{in anticipation} of a gravitational collapse that may take place a billion years from now! This phenomenon is concretely illustrated by collapse of a spherical, null fluid, described by the Vaidya solution to Einstein's equations, where one can explicitly see that the EH first forms and \emph{grows} in a flat (i.e. Minkowski) region of space-time, where nothing at all is happening. In the traditional descriptions of physical phenomena such teleology is regarded as spooky, and avoided studiously. For EHs, it occurs because the notion is unreasonably global; one needs to know the space-time structure all the way to the infinite future to say if admits an EH. In particular, in the simulations of compact binary mergers one cannot use EHs to locate either the progenitors or the final remnant as one numerically time-evolves the system since EHs can only be identified at the \emph{end} of the simulation, as an after thought. Given these serious limitations, a question naturally arises: Are there alternate, quasi-local horizons that can serve to characterize BHs without teleology? Can one achieve this and still maintain the attractive features of the EH mechanics? Advances over the past ~2 decades have shown that the answer is in the affirmative. We will focus on these developments. 

We first recall the notion of EHs, then introduce quasi-local horizons and discuss their mechanics. For brevity, all manifolds (and fields) are assumed to be smooth, with a metric of signature -,+,+,+. An arrow under a space-time index denotes the pull-back of that index to the horizon. We assume that matter satisfies a weak version of the dominant energy condition: for each future directed causal vector $t^a$, $-T_{ab} t^b$ is also future directed and causal \emph{at the horizon}. We will set the speed of light $c$ and Boltzmann's constant $k_{\rm B}$ to one.  Finally, unless otherwise stated, space-time is assumed to be 4-dimensional, and $g_{ab}$ is assumed to satisfy Einstein's equations with zero cosmological constant.

\section{Event Horizons}
\label{s2}

To introduce EHs, one starts with Penrose's conformal boundary $\scri$, representing future null infinity of asymptotically flat space-times. A \emph{Black-hole region} $\B$ of a space-time $(\M,g_{ab})$ is defined as $\B = \M \setminus I^{-}(\scri)$, where $I^-$ denotes `chronological past'.%
\footnote{For this notion to be physically useful, it is important that $\scrip$ be complete (Geroch and Horowitz 1978). With an incomplete $\scrip$, one would conclude that even Minkowski space has a black hole!}
The boundary $\partial\B$ of the black hole region is called the \emph{event horizon} and denoted by $\E$. Thus, $\E$ is the future boundary of the causal past of $\scri$. It therefore follows that $\E$ is a null 3-surface, ruled by null geodesics that are inextensible to the future. If the space-time is globally hyperbolic, an `instant of time'  is represented by a Cauchy surface $M$. The intersection of $\B$ with $M$ may have several disjoint components, each representing a black hole at that instant of time. If $M'$ is a Cauchy surface to the future of $M$, the number of disjoint components of $M'\cap\B$ in the causal future of $M\cap \B$ must be less than or equal to those of $M\cap \B$ (see Hawking \& Ellis 1973). Thus, black holes can merge but can not bifurcate. (By a time reversal, i.e. by replacing $\scri$ with $\mathscr{I}^-$ and $I^-$ with $I^+$, one can define a white hole region $\W$. However, here we will focus only on black holes.)

Early discussions of black hole mechanics focused on space-times that are stationary and axisymmetric. In these space-times, the EH $\E$ represents an equilibrium state of the BH. It is a Killing horizon, i.e., a linear combination 
\be K^a = t^a + \Omega\, \phi^a \ee
of the stationary Killing field $t^a$ and the rotational Killing field $\phi^a$ provides a canonical null normal to $\E$, where $t^a$ is normalized to be unit at infinity, $\phi^a$ is normalized so the affine length of its integral curves is
$2\pi$, and the constant $\Omega$ serves as the angular velocity of the horizon. It then follows that the \emph{surface gravity} $\k$ of the Killing normal $K^a$, defined by 
\be K^a \nabla_a K^b = \k\, K^b \, ,\ee
is constant on $\E$, even when $\E$ is distorted (e.g., by presence of matter rings). This is the analog of the zeroth law of thermodynamics that says that the temperature of a system in equilibrium is constant. If one passes from a given stationary axisymmetric space-time to one nearby, then the parameters of the BH change via 
\be \label{1law} \delta m = \f{\k}{8\pi G}\, \delta \a + \Omega\,
\delta J + \Phi\, \delta Q\ee
where $m, J$ denote the total mass and angular momentum of space-time, $a$ is the area of any 2-sphere cross-section of $E$ and $\Phi = A_a K^a$ is the electrostatic potential (Bardeen, Carter and Hawking 1973). The last two terms, $\Omega\,\delta J$ and $ \Phi\, \delta Q$, have the interpretation of `work' required to spin the black hole up by an amount $\delta J$ or to increase its charge by $\delta Q$. Therefore (\ref{1law}) has a striking resemblance to the first law, $\delta E = T \delta S + \delta W$, of thermodynamics if (as the zeroth law suggests) a multiple of $\kappa$ is identified with the temperature $T$, and a corresponding multiple of the horizon area $a$ with the entropy $S$.  Therefore, (\ref{1law}) and its generalizations discussed below are referred to as the \emph{first law of black hole mechanics.} A natural question now is whether there is an analog of the second law of thermodynamics. Using event horizons, Hawking showed that the answer is in the affirmative (see Hawking \& Ellis 1973). Let $(\M, g_{ab})$ admit an event horizon $\E$. Denote by $\l^a$ a geodesic null normal to $\E$. Its expansion is defined as $\theta_{(\l)} := q^{ab} \nabla_a \l_b$, where $q^{ab}$ is any inverse of the degenerate intrinsic metric $q_{ab}$ on $\E$, and determines the rate of change of the area-element of $\E$ along $\l^a$. Assuming that the null energy condition and Einstein's equations hold, the Raychaudhuri equation immediately implies that if $\theta_{(\l)}$ were to become negative somewhere it would become infinite within a finite affine parameter. Hawking showed that, if there is a globally hyperbolic region containing $I^-(\scri) \cup \E$ ---i.e., if there are no naked singularities--- this can not happen, whence $\theta{(\l)} \ge 0$ on $\E$. Therefore, if a cross-section $S_2$ of $\E$ is to the future of a cross section $S_1$, we must have $a_{S_2} \ge a_{S_1}$. Thus, in any (i.e., not necessarily infinitesimal) dynamical process, the change $\Delta a$ in the horizon area is always non-negative. This result is known as the \emph{second law of black hole mechanics.} As in the first law, the analog of entropy is the horizon area. Subsequently, Hawking's seminal discovery of black hole evaporation (Hawking 1974)
provided the precise relation between the laws of EH mechanics and thermodynamics: It is $\kappa\hbar /2\pi$ that plays the role of the thermodynamic temperature $T$ and $a/4G \hbar = a/4\lp^2$ that plays the role of entropy $S$. The factors of $\hbar$ --that are essential on dimensional grounds-- indicate that the resemblance of laws of EH mechanics with those of thermodynamics is an imprint left by quantum mechanics on classical physics.

In spite of this deep and fascinating feature, the analogy between EH mechanics and thermodynamics has some important limitations. Let us begin with the first law. In thermodynamics, one only assumes that the physical system under consideration is in equilibrium, while in the discussion of the first law of EH mechanics one assumes that the entire space-time --including the far away matter, for example-- is stationary, not just the EH. Secondly, while all quantities that enter the statement of the first law of thermodynamics refer just to the system under consideration, now $m, J$ refer not just to the BH but to everything contained in the space-time, since they are calculated at spatial infinity. Finally, in the second law of thermodynamics, change in entropy of a given system is directly related to physical processes occurring in its neighborhood. For event horizons, as we saw, this is not the case; area of cross-sections of $\E$ can increase for teleological reasons, in anticipation of a physical process that would occur in, say, a billion years to the future, even when nothing at all is falling across $\E$ for all these years! 

Quasi-local horizons were introduced to improve on this situation. \emph{Isolated horizons}, dented by $\IH$ in the literature, provide a quasi-local description of situations in which it is only the BH that is in equilibrium; there may be dynamical processes away from it. A Dynamical Horizon $\DH$ represents BHs whose horizon structure is changing in time due to the full non-linear dynamics of GR. Change in its area is due to instantaneous fluxes across $\DH$; there is no teleology.

\section{Isolated Horizons: Local Equilibrium}
\label{s3} 

The key idea here is drop the requirement that the entire space-time should admit a stationary Killing field and ask only that the intrinsic horizon geometry be time-independent. Consider a null, 3-dimensional sub-manifold $\IH$ with topology  $\S^2\times \R$ in a space-time $(\M,g_{ab})$, and denote future pointing normals to $\IH$ by $\l^a$. The pull-back $q_{ab} := g_{\pback{ab}}$ of the space-time metric to $\IH$ is the intrinsic, degenerate `metric' of $\IH$ with signature 0,+,+. (Recall that an arrow under a space-time index denotes the pull-back of that index to the horizon.) The first condition is that $q_{ab}$ be `time-independent', i.e. $\Lie_{\ell}\, q_{ab} =0$ on $\IH$. This condition implies that the area of any 2-sphere cross-section of $\IH$ is constant. Therefore, such sub-manifolds $\IH$ are referred to as \emph{non-expanding horizons} (NEHs). One can show that every NEH inherits a natural derivative operator $D$, obtained by the restricting of the space-time derivative operator $\nabla$ to $\IH$. While $D$ is compatible with $q_{ab}$, i.e. $D_a q_{bc} =0$, it is not uniquely determined by this property because $q_{ab}$ is degenerate. Thus, $D$ has extra information, not contained in $q_{ab}$. The pair $(q_{ab}, D)$ is said to determine the \emph{intrinsic geometry} of the null surface $\IH$. This notion leads to a natural notion of a horizon in local equilibrium.
\vskip0.1cm

\ni {\sl Definition 1:} A NEH $\IH$ is said to be \emph{isolated horizon} (IH) if it admits a null normal $\l^a$ such that:\,\,
$\Lie_{\l}\, q_{ab} = 0$ and $[\Lie_{\l},\, D] =0$ on $\IH$.\vskip0.1cm
\ni On can show that, generically, this null normal field $\l^a$ is unique up to rescalings by positive \emph{constants} (Ashtekar, Beetle, Lewandowski 2002). 

Note that the definition refers just to $\IH$; in particular, $(\M, g_{ab})$ is not required to be asymptotically flat and there is no longer any teleological feature. Since $\IH$ is null and $\Lie_{\l} q_{ab} =0$, the area of \emph{any} of its cross sections is the same, denoted by $a_\IH$. As one would expect, one can show that there is no flux of gravitational radiation or matter across $\IH$. This captures the idea that the black hole itself is in equilibrium. Thus, \emph{Definition 1} extracts from the notion of a Killing horizon just a `tiny part' that refers only to the intrinsic geometry of $\IH$. As a result, every Killing horizon is, in particular, an IH, whence the EH of any stationary BH is an IH. However, IHs are more general. For example, the multi-black hole solution of Kastor and Traschen admits IHs which are not Killing horizons. Also,  a space-time with an IH $\IH$ can admit gravitational radiation and dynamical matter fields away from $\IH$. In fact, as a family of Robinson-Trautman space-times illustrates, gravitational radiation could even be present arbitrarily close to $\IH$. Because of these possibilities, the transition from EHs of stationary space-times to IHs represents a significant generalization of black hole mechanics.%
\footnote{In fact the derivation of the zeroth and the first law
requires slightly weaker assumptions, encoded in the notion of a
`weakly IH' (Ashtekar et al 2000, 2001).}

An immediate consequence of the requirement $\Lie_{\l} q_{ab} =0$
is that there exists a 1-form $\omega_a$ on $\IH$ such that $D_a
\l^b = \omega_a \l^b$. Following the definition of $\k$ on a
Killing horizon, the \emph{surface gravity} $\k_{(\l)}$ of $(\IH,
\l)$ is defined as $\k_{(\l)} = \omega_a\l^a$. Again, under $\l^a
\rightarrow c\l^a$, we have $\k_{(c\l)} = c\k_{\l}$. Together with
Einstein's equations, the two conditions of \emph{Definition 1} imply\,
$\Lie_{\ell}\, \omega_a =0$ and\, $\l^a D_{[a}\omega_{b]} =0$. The
Cartan identity relating the Lie and exterior derivative now
yields
\be D_a (\omega_b \l^b) \equiv D_a \k_{(\l)} = 0\, .\ee
\emph{Thus, surface gravity is constant on every isolated
horizon}. This is the zeroth law, extended to horizons
representing local equilibrium. In presence of an electromagnetic
field, \emph{Definition 1} and the field equations imply: $\Lie_{\l}\,
F_{\pback{ab}} =0$ and $\l^a F_{\pback{ab}} =0$. The first of
these equations implies that one can always choose a gauge in
which $\Lie_{\l} A_{\pback a} =0$. By Cartan identity it then
follows that the electrostatic potential $\Phi_{(\l)} := A_a \l^a$
is constant on the horizon. This is the Maxwell analog of the
zeroth law.

In this setting, the first law is derived using a Hamiltonian
framework (Ashtekar et al 2000, 2001). For concreteness and simplicity, let us
assume that we are in the asymptotically flat situation and the
only matter field present is electromagnetic. One begins by
restricting oneself to horizon geometries such that $\IH$ admits a
rotational vector field $\varphi^a$ satisfying%
\footnote{For black hole mechanics, in fact it suffices to assume
only that $\Lie_{\varphi} \epsilon_{ab} =0$ where $\epsilon_{ab}$
is the intrinsic area 2-form on $\IH$. The same is true on
dynamical horizons discussed in the next section.}
$\Lie_{\varphi} q_{ab} =0$. One then constructs a phase space $\G$
of gravitational and matter fields such that:  i) the space-time manifold $\M$ admits an
internal boundary $\IH$ which is an IH; and, ii) all
fields satisfy asymptotically flat boundary conditions at
infinity. Note that the horizon geometry is allowed to vary from
one phase space point to another; the pair $(q_{ab}, D)$ induced
on $\IH$ by the space-time metric has to satisfy only \emph{Definition 1}
and the condition $\Lie_{\varphi} q_{ab} =0$.

Let us begin with angular momentum. Fix a vector field  $\phi^a$
on $\M$ which coincides with the fixed $\varphi^a$ on $\IH$ and is
an asymptotic rotational symmetry at infinity. (Note that $\phi^a$
is not restricted in any way in the bulk.) Lie derivatives of
gravitational and matter fields along $\phi^a$ define a vector
field ${\X}(\phi)$ on $\G$. One shows that it is an infinitesimal
canonical transformation, i.e., satisfies $\Lie_{\X(\phi)}\, \O
=0$ where $\O$ is the symplectic structure on $\G$. The
Hamiltonian $\H(\phi)$ generating this canonical transformation is
given by:
\be \H(\phi) = J_\IH^{(\phi)}  - J_\infty^{(\phi)} \quad {\rm
where} \quad J_\IH^{(\phi)}  = -\f{1}{8\pi G}\, \oint_\S (\omega_a
\varphi^a)\, \epsilon - \f{1}{4\pi}\, \oint_\S (A_a\varphi^a) \,
{}^\star{F} \ee
where $J_\infty^{(\phi)}$ is the total angular momentum at spatial infinity, $\S$ is any cross-section of $\IH$ and $\epsilon$ the area element thereon. The term $J_\IH^{(\phi)}$ depends \emph{only} on $\varphi$ and the geometry of $\IH$, and is independent of the choice of the 2-sphere $\S$ made in its evaluation. It is interpreted as the \emph{horizon angular momentum}. It has numerous properties that support this interpretation. In particular, it yields the standard angular momentum expression in Kerr-Newman space-times.

To define horizon energy, one has to introduce a `time-translation' vector field $t^a$. On $\IH$, it must be a symmetry of $q_{ab}$. Since $\l^a$ and $\varphi^a$ are both horizon symmetries, \, one sets $t^a = c \l^a + \Omega \varphi^a$\, on $\IH$, for some constants $c$ and $\Omega$. However, unlike $\phi^a$, the restriction of $t^a$ to $\IH$ can not be fixed once and for all but must be allowed to vary from one phase space point to another. In particular, on physical grounds, one expects $\Omega$ to be zero at a phase space point representing a non-rotating black hole, but non-zero at a point representing a rotating black hole. This freedom in the boundary value of $t^a$ introduces a qualitatively new element. The vector field $\X(t)$ on $\G$ defined by the Lie derivatives of gravitational and matter fields does not, in general, satisfy $\Lie_{\X(t)}\, \O =0$; it need not be an infinitesimal canonical transformation. The necessary and sufficient condition is that \, $ ( \k_{(c\l)}/8\pi G)\delta a_{{}_\IH} + \Omega \delta J_\IH + \Phi_{(c\l)} \delta Q_\IH$\, be an exact variation. That is, $\X(t)$ generates a Hamiltonian flow if and only if there exists a function $E^{(t)}_\IH$ on $\G$ such that
\be \label{1law3}\delta E^{(t)}_\IH = \f{\k_{(c\l)}}{8\pi G}
\delta a_{{}_\IH} + \Omega \delta J_\IH + \Phi_{(c\l)} \delta Q_\IH \ee
This is precisely the first law! Thus, the framework provides a
deeper insight into the origin of the first law: It is the
necessary and sufficient condition for the evolution generated by
$t^a$ to be Hamiltonian. Eq. (\ref{1law3}) is a genuine restriction on
the choice of phase space functions $c$ and $\Omega$, i.e., of
restrictions to $\IH$ of evolution fields $t^a$. It is easy to
verify that $\M$ admits many such vector fields. Given one, the
Hamiltonian $H(t)$ generating the time evolution along $t^a$ takes
the form
\be \H(t)\, =\, E^{(t)}_\infty - E^{(t)}_\IH\, , \ee
where the expression of $E^{(t)}_\IH$ refers only to fields on $\IH$, 
re-enforcing the interpretation of $E^{(t)}_\IH$ as the horizon
energy.

Thus, in general, there is a multitude of first laws, one for each
vector field $t^a$, the evolution along which preserves the
symplectic structure. In the Einstein-Maxwell theory, given
\emph{any} phase space point, one can choose a canonical boundary
value $t_o^a$ by exploiting the black hole uniqueness theorem. $E^{(t_o)}_\IH$
is then called the horizon mass and denoted simply by $m_\IH$. In
the Kerr-Newman family, $\H(t_o)$ vanishes and $m_\IH$ coincides
with the ADM mass $m_\infty$. Similarly, if $\phi^a$ is chosen to
be a global rotational Killing field, $J_\IH^{(\phi)} $ equals
$J_\infty^{(\phi)}$. However, in more general space-times where
there is matter field or gravitational radiation outside $\IH$,
these equalities do not hold. The $m_{{}_\IH}$ and $J_\IH$  that enter the IH first law (\ref{1law3}) represent
quantities associated with the \emph{horizon alone}, while $m$ and $J$ that feature in the EH first law (\ref{1law}) are the \emph{total} mass angular momentum in the space-time, measured at spatial infinity, including contributions from matter fields that may be present in the exterior region. 

When the uniqueness theorem fails, as for example in the Einstein-Yang-Mills-Higgs theory, the infinite family of first laws continue to hold, but the horizon mass $m_\IH$ becomes ambiguous. Interestingly, these ambiguities can be exploited to relate properties of hairy black holes with those of the corresponding solitons. (For a summary, see Ashtekar \& Krishnan (2004).)

\section{Dynamical situations}
\label{s4}

It is tempting to ask if there is a local physical process directly responsible for the growth of horizon area (as there is for change of entropy of a closed thermodynamical system). As we saw, the answer is in the negative for EHs since they can grow in a flat portion of space-time. However, one can introduce quasi-local horizons also in dynamical situations and obtain the desired result (Ashtekar \& Krishnan 2003). These constructions are motivated in part by earlier ideas introduced by Hayward (1994). 

Since the subject of black hole dynamics is rich and diverse, one cannot incorporate all its aspects in a single, concise treatment. In this article we focus on classical aspects. For a summary of the role played by quasi-local horizons in quantum dynamics of black holes, see Ashtekar 2020. \\ 
 \ni \textsl{Definition 2:} A 3-dimensional space-like sub-manifold $\DH$ of $(\M,g_{ab})$ is said to be a \emph{dynamical  horizon} (DH) if it admits a foliation by compact 2-manifolds $\S$ (without boundary) such that the expansion $\theta_{(\l)}$ of  one (future directed) null normal field $\l^a$ to $\S$ vanishes and the expansion of the other (future directed) null normal  field, $n^a$ is negative.\vskip0.1cm

One can show that this foliation of $\DH$ is unique (Ashtekar \& Galloway 2005) and that $\S$ has the topology of a 2-sphere (except in some degenerate and physically over-restrictive cases when it has the topology of a 2-torus). Each leaf $\S$ is a marginally trapped surface (MTS) and will be referred to as a \emph{cut} of $\DH$. Unlike event horizons $\E$, DHs $\DH$ are locally defined and do not display any teleological feature. In particular, they \emph{cannot} lie in a flat portion of space-time. DHs arise in numerical simulations of binary black hole mergers as world tubes of MTSs. If a MTS is located on a Cauchy slice during evolution, generically it persists (Andersson et al 2009). In the early days, numerical simulations focused on outermost MTSs on the 3+1 foliation they used. This led to a general belief that MTSs jump abruptly when a common horizon is formed. More recent high precision numerical work has shown that this conclusion is incorrect. In fact the world tubes of MTSs associated with individual black holes persist and join on continuously to the final, common world tube representing the remnant (Pook-Kolb et al 2019, 2021). During this process, the world tubes are not everywhere space-like, but their early and late portions are. Thus, the individual world tubes of MTSs representing the progenitors while they are well separated as well as the final portion of the world tube representing the remnant are DHs. As the remnant settles down, its DH approaches the EH. During the  dynamical phase, $\DH$ lies inside $\E$.

\emph{Definition 2} immediately implies that the area of cuts of  $\DH$ increases monotonically along the `outward direction' defined by the projection of $\l^a$ on $\DH$. More importantly, a detailed analysis shows that this  change is directly related to the flux of energy falling across $\DH$. Denote by $R$ the area radius of the MTS cuts $\S$ of $\DH$, so that $a_\S = 4\pi R^2$. Let $N$ denote the norm of $\partial_a R$ and $\DH_{1,2}$ the portion of $\DH$ bounded by two cross-sections $\S_1$ and $\S_2$. The appropriate notion of energy flux turns out to be associated with the vector field $N\l^a$ where $\l^a$ is normalized such that its projection on $\DH$ is the unit normal $\hat{r}^a$ to the cuts $\S$. In the generic and physically interesting case when $\S$ is a 2-sphere, the Gauss and the Codazzi (i.e. constraint) equations imply:
\be \label{area-increase} \f{1}{2G}\, (R_2 - R_1) =  \int_{\DH_{{}_{1,2}}} T_{ab}\, N\l^a
\hat{\tau}^b d^3V + \f{1}{16\pi G}\, \int_{\DH_{{}_{1,2}}} N\, \left(
\sigma_{ab}\sigma^{ab} + 2 \zeta_a\zeta^a\right)\, d^3V . \ee
Here $\hat{\tau}^a$ is the unit normal to $\DH$,\,
$\sigma^{ab}$,\, is the shear of $\l^a$ (i.e., the trace-free part
of $q^{am}q^{bm} \nabla_m \l_n$)\, and $\zeta^a = q^{ab}\hat{r}^c
\nabla_c \l_b$, where $q^{ab}$ is the projector onto the tangent
space of the cuts $\S$. The first integral on the right can be
directly interpreted as the flux across $\DH_{1,2} $ of
matter-energy (relative to the causal vector field $N \l^a$). The second
term is purely geometric and is interpreted as the flux of energy
carried by gravitational waves across $\DH_{1,2} $. It has several
properties which support this interpretation. Thus, not only does
the second law of black hole mechanics hold for a dynamical
horizon $\DH$, but the `cause' of the increase in the area can be
directly traced to physical processes happening near $\DH$.

Another natural question is whether the first law (\ref{1law3})
can be generalized to fully dynamical situations, where $\delta$
is replaced by a finite transition. Again, the answer is in the
affirmative. We will outline the idea for the case when there are
no gauge fields on $\DH$. As with IHs, to have a
well-defined notion of angular momentum, let us suppose that the
intrinsic 3-metric on $\DH$ admits a rotational Killing field
$\varphi$. Then, the angular momentum
associated with any cut $\S$ is given by%
\begin{equation} \label{jdynamic1}J_{\S}^{(\varphi)} =
-\frac{1}{8\pi G} \oint_{\S} K_{ab} \varphi^a\hat{r}^{\,b} \,
d^2V\, \equiv \frac{1}{8\pi G}\oint_{\S} j^{(\varphi)} d^2V\, ,
\end{equation}
where $K_{ab}$ is the extrinsic curvature of $\DH$ in $(\M,
g_{ab})$ and $j^{(\varphi)}$ is interpreted as `the angular
momentum density'. Now, in the Kerr family, the mass,
surface-gravity and the angular velocity can be unambiguously
expressed as well-defined functions $\bar{m}(a,J)$, $\bar\k(a, J)$
and $\bar\Omega(a, J)$ of the horizon area $a$ and angular
momentum $J$. The idea is to use these expressions to associate
mass, surface gravity and angular velocity with each cut of $\DH$. Intuitively, this corresponds to identifying the geometrical fields on each MTS of $\DH$ as providing the geometry of a Kerr IH, so that the evolution along $\DH$ is seen as providing a 1-parameter family of Kerr IHs. Then, a surprising result is that the difference between the
horizon masses associated with cuts $\S_1$ and $\S_2$ can be
expressed as the integral of a \emph{locally defined} flux across
the portion $\DH_{1,2} $ of $\DH$ bounded by $\DH_1$ and $\DH_2$:
\begin{eqnarray}
\label{1law4} \bar{m}_2 - \bar{m}_1 &=& \f{1}{8\pi G}\,
\int_{\DH_{1,2} } \bar\kappa da\nonumber\\
&+& \frac{1}{8\pi G} \biggl\{ \oint_{\S_2}\bar\Omega
j^\varphi\,d^2V -\oint_{\S_1} \bar\Omega j^\varphi\,d^2V \biggr. -
\biggl.\int_{\bar\Omega_1}^{\bar\Omega_2} d\bar\Omega \oint_\S
j^\varphi\,d^2V \biggr\} \end{eqnarray}
If the cuts $\S_2$ and $\S_1$ are only infinitesimally separated,
this expression reduces precisely to the standard first law
involving infinitesimal variations. Therefore, (\ref{1law4}) is
\emph{an integral generalization of the first law}.

\section{Multipole moments and tomography}
\label{s5}

IHs and DHs have provided valuable insights on a number of issues that cannot be analyzed using EHs because of their teleological nature. For example, the IHs framework provides natural boundary conditions (at the inner boundaries) for the initial value problem in which the black holes can be regarded as being in quasi-equilibrium, e.g., when they are sufficiently far away from each other (Jaramillo 2011). Another example is the determination of the mass and the spin of the progenitors as well as of the final remnant in a binary black hole merger using procedures outlined in section \ref{s3}.  These parameters in turn lead to a notion of `binding energy' between the black holes that is useful in locating `quasi-circular' orbits. (For details, other applications, and further references, see Ashtekar \& Krishnan 2004.) Form a conceptual and mathematical viewpoint, a more powerful tool provided by quasi-local horizons is the invariant characterization of the horizon geometry via multipole moments. Properties of multipoles implied by Einstein's equations provide us with a richer body of laws of horizon mechanics.
\vskip0.1cm

\textbf{Multipole Moments:$\!$ IHs.}\, Black hole uniqueness theorems for vacuum (and electrovac) solutions assure us that the horizon geometry of isolated stationary black holes is well-described by that of the Kerr(-Newman) EH. However, astrophysical black holes are rarely completely isolated --there can be matter rings or another black hole that can perturb the horizon geometry. The possibility of this rich structure is well-captured by IHs, where the intrinsic metric $q_{ab}$ and the derivative operator $D$ are sensitive to these external influences. The challenge is to provide an invariant characterization of these distortions in the shape and spin structure of the horizon. This is provided by two sets of numbers $I_{\ell,m}$ representing the shape multipoles, and $L_{\ell,m}$ representing the spin multipoles. One may be given two space-times with IHs, presented in different coordinates, representing, for example, progenitor black holes in the distant past constituting a binary, or, remnants after the merger in the distant future. Do the two IHs represent same physics? A priori, this question seems difficult to address. Multipole moments provide a natural avenue to answer this question. In each space-time one can separately compute the multipoles using the coordinates and other set up used to describe its IHs. Then IHs are physically the same if and only if the $I_{\ell,m}$ and $L_{\ell,m}$ all agree. 

Recall that there is no flux of matter across an IH. However, there may be time independent (Maxwell) fields. For simplicity, let us first suppose that there are no matter fields on the given IH itself (although there may be, e.g., matter rings, outside). Then, Einstein's equations imply that the IH geometry is completely determined by the null-normal $\ell^a$, a number $\kappa_{(\ell)}$, and two fields defined intrinsically on a 2-sphere cross-section $\S$ of $\IH$: a metric $\t{q}_{ab}$ of signature +\,+, and a 1-form $\t{\omega}_a$. (As one might expect, these fields are just pull-backs to $\S$ of the metric $q_{ab}$ and the rotation 1-form $\omega_a$ on $\IH$, and $\kappa_{(\ell)}$ is the surface gravity of $\ell^a$.) Since $\S$ has the topology of a 2-sphere, the invariant information in $\t{q}_{ab}$ is encoded in its scalar curvature $\t{R}$. The invariant information in $\t\omega_a$ is contained in its curl because $\t\omega_a$ changes by the addition of a gradient when one changes the initial cross-section $\t{S}$. Interestingly, on an IH, $\t{R}$ is proportional to the real part of the Newman-Penrose component $\Psi_2 $ of Weyl curvature, and the curl of $\t\omega_a$, in the imaginary part. Thus, the invariant information in the free data of the IH geometry is neatly encoded in $\Psi_2$, which is unambiguously defined because we have a canonical null vector field $\ell^a$ and the `radiative' field on $\IH$ ${}^\star C^{abcd}\l_b\l_d$ --or $\Psi_0$ and $\Psi_1$-- vanishes.

The idea is to encode $\Psi_2$ in two sets of numbers, $I_{\ell,m}$, and $L_{\ell,m}$, using invariantly defined spherical harmonics. This step can be carried out different ways. We will discuss the one that has been used most extensively so far. As in the discussion of the first law, one assumes that the metric $q_{ab}$ on $\IH$ is axisymmetric.%
\footnote{For a discussion of multipoles on a generic IH that are not axisymmetric --in fact on a generic NEH-- see Ashtekar, et al (2022a).}
Then, one can introduce a canonical \emph{round} 2-sphere metric $\qo_{ab}$ which shares the rotational Killing field, as well as the area 2-form with $q_{ab}$. This construction is invariant in that it does not require introduction of any extra structure. One then uses the spherical harmonics $\Yo_{\ell,m}$ of $\qo_{ab}$ and sets:
\be \label{mm1} I_{\ell,m} + i L_{\ell,m} \, :=\, - \oint_{\S} \Psi_2 \, \Yo_{\ell,m} \rmd^2 \t{V} \ee
where $\rmd^2 \t{V}$ is the volume element induced on $\S$ by $\t{q}_{ab}$ (as well as by $\qo_{ab}$). One can show that the right side is independent of the choice of the cross-section $\S$; thus the shape and spin multipoles are properties of $\Delta$ alone. This set characterizes the IH geometry in the following sense: Given an IH, one can reconstruct its geometry $(q_{ab}, \, D)$, starting from its multipoles $I_{\ell,m},\,  L_{\ell,m}$ (given the vector field $\ell^a$ and two numbers: $\kappa_{\ell}$ and $a_{{}_\S}$ on a bare manifold $\S^2 \times \R$).

These \emph{geometrical} multipole moments are dimensionless. The \emph{mass and angular momentum }moments, $M_{\l, m}$ and $J_{\l, m}$, that are of more direct physical interest can be obtained simply by rescaling the $I_{\ell,m}$ and $L_{\ell,m}$ by appropriate dimensionfull factors.
The proportionality factors are determined by analogy to electrodynamics where charge and current moments suffice to determine the source. In electrodynamics, one uses charge and current densities to define the moments. For the IHs under consideration, there are no matter fields on $\IH$; the required `densities' arise from the horizon geometry itself. A comparison between the expressions of angular momentum and the mass of IHs and the integral on the right side of (\ref{mm1}) yields the required expressions of the `effective mass and angular momentum densities'. They only involve powers of $m_{{}_\IH}$, and the area radius $R_{\IH}$ of the IH which then determine the proportionality factors between $I_{\l,m}$ and $M_{\l, m}$, and between $L_{\l,m}$ and $J_{\l, m}$.

These mass and angular momentum multipoles have several attractive properties. First, the mass dipole and the angular momentum monopole always vanish. The mass monopole equals $m_{{}_\IH}$, and the angular momentum dipole equals the $J_\IH^{(\phi)}$  discussed in section \ref{s3}. Second, in vacuum, stationary space-times the mass monopole equals the ADM mass $m_{{}_{\infty}}$ at spatial infinity. Third, in static space-times all angular momentum multipoles vanish.  Finally, in vacuum axisymmetric space-times, the angular momentum dipole $J_{1,0}$ equals $J_\infty$. In particular, then, for the Kerr family, the mass monopole and angular momentum dipole are the expected ones, and the discrete reflection symmetry of the spacetime metric implies that all odd mass multipoles and all even angular momentum multipoles vanish. However, that there is a subtlety about the higher multipoles whose values are not determined by space-time Killing fields: The multipole moments defined here are the `source multipoles' that refer to the horizon, and are thus distinct from the `field multipoles' defined at spatial infinity. In Newtonian gravity the two sets agree. However, in general relativity, the gravitational field outside the horizon also contributes to the field multipoles since they are extracted from the gravitational field \emph{at infinity}. The discrepancy increases with the Kerr parameter $a$. In the extremal limit, $a\to 1$, the field mass-quadruple moment is smaller than the horizon one by a factor of 1.4 while the field angular momentum octupole moment is greater than the horizon one by a factor of 1.14. These differences are conceptually important for 
certain applications such as the `self-force' calculations used in the study of extremal mass ratio binaries, where it is the source or the horizon multipoles that are relevant.

These considerations have been extended to include Maxwell fields at $\IH$. Then, in addition to the mass and angular momentum moments, one also has the charge and current multipoles. 
(For further details in IH multipoles, see Ashtekar et al (2004).)\vskip0.1cm

\textbf{Multipole Moments:$\!$ DHs.}\, Let us now turn to dynamical situations and assume that there are no matter fields on the horizon itself. A black hole may be formed by a gravitational collapse or as a result of a compact binary coalescence. In these processes, a dynamical horizon forms and eventually settles down to a Kerr IH in the distant future. However, as numerical simulations show, initially the shape and spin structure of the DH is quite complicated, and changes rapidly in time. Can one provide an invariant description of the fascinating dynamics in the fully non-linear regime of general relativity? The answer is in the affirmative. One can capture the information in the geometry of each $v={\rm const}$ cut $\S$ of the DH in time-dependent multipoles, i.e., functions $I_{\l,m}(v),\, L_{\l,m}(v)$ of time. This 1-parameter family of shape and spin multipoles provides an invariant characterization of the horizon dynamics. (For details, see Ashtekar et al 2013).     

As in section \ref{s4}, one can think of each cut $\S$ of $\DH$ as a cross-section of an isolated horizon $\IH$. The shape is again invariantly encoded in the scalar curvature $\t{R}(v)$ of the 2-metric $\t{q}_{ab}(v)$ on $\S$. However, since $\t{q}_{ab}(v)$ is now time dependent, $\t{R}(v)$ is no longer proportional to ${\rm Re} \Psi_2(v)$. As before, the spin structure is encoded in the `rotation' 1-form, now defined as  $\t\omega_a (v):= -n^c\,\t{q}_a{}^b\,\nabla_b \ell_c (v)$ on $\S$. (The expression on the right side yields the rotational one from $\t\omega_a$ also on an IH.) However, now there is no gauge freedom in $\t\omega_a(v)$ because it is tied to a specific cut, i.e., an MTS that has direct physical meaning. Finally, because of time dependence, its curl is no longer given by ${\rm Im} \Psi_2(v)$. Therefore, in the definition of multipoles, the seed function that replaces $\Psi_2$ in  (\ref{mm1}) now given by $\Phi(v) := - \f{1}{4}\,\t{R}(v) + \f{i}{2} \t\epsilon^{ab} \t{D}_a \t{\omega}_b (v)$ (which becomes $v$ independent on an IH and reduces to $ \Psi_2$). This function encodes the information in the spin and mass structure of the DH on each cut $\S$.

Recall that one also needs spherical harmonics to define the multipoles. Now, in the dynamical phase of a collapse or a merger, $\t{q}_{ab}(v)$ is far from being axisymmetric, whence the procedure to extract a canonical round metric used on IHs now fails. However, in the asymptotic future the DH tends to the Kerr IH which \emph{is} axisymmetric. Therefore, a natural strategy would be to introduce spherical harmonics $\Yo_{\l,m}$ in the asymptotic future and Lie-drag them to the rest of $\DH$ along a suitable vector field $X^a$ on $\IH$. However, for this procedure to be physically meaningful, the vector field $X^a$ has to satisfy several constraints. For example: (i) $X^a$ has to be constructed by a covariant procedure that does not require any additional structure; (ii) the flow it generates has to preserve the foliation of $\IH$ by MTSs $\S$;  (iii) if the DH happens to be axisymmetric, then the flow has to preserve the rotational Killing vector so that the $\Yo_{\l,m}$ obtained by the dragging procedure agree with those constructed on each leaf; etc. It turns out that, thanks to  these constraints, $X^a$ is unique up to a rescaling $X^a \to f(v) X^a$ for some positive function $f(v)$, and this small freedom does not affect the $\Yo_{\l,m}$ it defines on all of $\DH$, and hence the multipoles they define. 
The time-dependent multipoles on a DH are given by:
\be I_{\ell,m}(v) + i L_{\ell,m}(v) \, :=\, -\oint_{\S_v} \Phi(v) \, \Yo_{\ell,m}\, \rmd^2 \t{V}. \ee
Einstein's equations relate the change in the multipoles between two cuts, $v\!=\!v_1$ and $v\!=\!v_2$, to fluxes of geometric fields across the portion $\DH_{1,2}$ of $\DH$ bounded by the two cuts. One thus obtains two infinite family of laws of DH mechanics, each labeled by $\ell,m$, that supplement the the three laws discussed in sections \ref{s2} and \ref{s3} by carrying  
much more detailed information about horizon dynamics. 

These moments have been calculated using numerical simulations of binary black hole mergers. On the one hand, the laws governing the change of multipoles dictated by Einstein's equations serve as a tool to check the accuracy of simulations. On the other hand, the simulations serve to 
quantify the appearance of distortions in the shape and spin structures of individual horizons of the progenitors as they approach each other, as well as the disappearance of these distortions as the final remnant settles down to the Kerr IH. Let us consider the post-merger phase first and discuss the progenitors in the next sub-section.

Gupta et al (2018), for example, provide detailed information on the ($\l$-dependent) decay of the first few multipoles as the remnant DH asymptotes to the Kerr IH. It is often said that the complicated multipolar structure --or the `hair'-- at the formation of the common DH is `radiated away' to infinity as the DH settles down to the Kerr IH. However, this description of the process is incorrect; no radiation can propagate from a DH (or an EH) to infinity because of the causal structure of space-time. Rather, it is the radiation falling \emph{into the DH} that irons out the irregularities so that the final multipoles are completely characterized just by the two Kerr parameters, $M$ and $a$. Now, as discussed below Eq. (\ref{area-increase}), $|\sigma_{ab}|^2_\DH$ can be interpreted as the energy density associated with gravitational waves falling \emph{into} $\DH$ that is directly responsible for the increase in area of the cuts $\S$. Detailed analysis has brought out correlations between the angular distribution of this infalling flux and the damping of the differences between the DH multipoles and their final Kerr values (Pook-Kolb et al 2020). Interestingly, the damping in the strong field regime is also correlated with the so-called quasi-normal damping of the waveform registered at $\scrip$ (discussed below). Thus, the mechanics of multipole moments provide sharp tools to probe rich interplay between geometry and physics in the strong field, dynamical regime of general relativity that had not been explored before. Note that one cannot carry out such investigations using EHs. For, one cannot define shape and spin multipoles on EHs because in the strongly dynamical regime, EHs fail to be smooth sub-manifolds --they have creases, corners and caustics (Gadioux \& Reall 2023). Even on portions that may be smooth at late times, EHs do not have a natural foliation that was crucially used to define multipoles on DHs. \vskip0.1cm

\textbf{Gravitational wave tomography:} As we noted above, there is no causal propagation between DHs $\DH$ (or EHs) and $\scrip$. Nonetheless, there are unforeseen correlations between the horizon dynamics and the strain measured by gravitational wave detectors far away!  As the term `tomography' suggests, even though DHs $\DH$ are not `visible' to asymptotic observers, thanks to Einstein's equations, one can create a movie of the time evolution of the shape and spin structure of $\DH$ using gravitational wave detectors in asymptotic regions. We will conclude with a discussion of the interplay between dynamics in the strong field and in the asymptotic regimes, suggested in Ashtekar \& Krishnan (2004).

Over the years, a number of simulations have provided concrete evidence for these fascinating correlations (for a summary of the early work, see Jaramillo 2012). Let us begin with the inspiral phase of the binary black hole coalescence when the two progenitors are relatively close, but before they merge. Then the waveform at $\scrip$ is oscillatory with increasing amplitude and frequency, characterized by a `chirp'. The energy flux at infinity is encoded in the  so-called Bond-news --the shear of the null normal $n^a$ to $\scrip$. As noted above, the energy-flux across the DH of each progenitor is encoded in the shear of the null normal $\ell^a$ to the cuts $\S(v)$ of $\DH$. These fields have been computed for mergers in which the initial black holes are non-spinning (Prasad et al 2020). Shear of the DHs of each progenitor also exhibits a chirp and furthermore all three shears are correlated, once the initial phases and times are matched. 

There have also been a number of simulations that focus on the time evolution of geometry and the multipolar structure of the DH of the remnant (see, in particular, Gupta et al 2018, Prasad et al 2022, Chen etal 2022). Together, they strongly hint that there is a detailed relation between the waveform extracted in the asymptotic region at $\scrip$, and the evolution of multipoles in the strong field region of $\DH$. An analytical understanding of this interplay is beginning to emerge using $I_{\ell,m}$ and $L_{\ell,m}$ at the horizon, and two infinite sets of observables at $\scrip$ --the Bondi-Sachs supermomenta $Q_{\ell,m}(u)$ and their `magnetic' (or, Newman-Unti-Tamburino) analogs $Q^\star_{\ell,m}(u)$-- each of which is also labelled by $\ell, m$. One uses the fact that, soon after the merger, the common DH is well approximated by a precisely defined `perturbed Kerr IH' (Ashtekar et al 2022b). Numerical simulations confirm that the perturbations are linear combinations of `quasi-normal modes'. These modes are ingoing at the horizon and outgoing at infinity and have complex frequencies that are completely determined by the parameters $M, a$ labelling the final Kerr IH. In this quasi-normal regime, using Einstein's equations one can show that there is a direct relation between the time-rate of change of $I_{\ell,m}$ and $L_{\ell,m}$ at $\H$, and and that of $Q_{\ell,m}$ and $Q^\star_{\ell,m}$ at $\scrip$. The waveform at $\scrip$ determines the rate of change of $Q_{\ell,m}$ and $Q^\star_{\ell,m}$ and the relation determines the horizon dynamics. Thus, one has a detailed tomography map in the quasi-normal regime. 

Finally, one can compare the exact time-dependent multipoles  $I_{\ell,m}(v)$ and $L_{\ell,m}(v)$ provided by numerical simulations with their final asymptotic Kerr values that correspond to the $v \to \infty$ limit. How far in the past can the difference be well approximated by the quasi-normal corrections to the Kerr multipoles, calculated \emph{using the waveform} by the procedure given above? For a given desired accuracy (determined by the detector sensitivity), the approximation could fail and nonlinear corrections may become important near the merger (Khera et al 2023). The multipole moment considerations provide a sharp criterion determine the domain on which the quasi-normal approximation is viable on entire exterior region of space-time. This tool is likely to play an important role in the ongoing debate (see, e.g., Giesler et al 2019, Baibhav et al 2023) on the domain of validity of the quasi-normal domain, already within general relativity.

\section{Discussion}
\label{s6}

Let us conclude with general perspectives. On the whole, in the passage from event horizons in stationary space-times to isolated horizons, and then to dynamical horizons, one considers increasingly more realistic situations. The laws of horizon mechanics follow from the specific definitions of these horizons and Einstein's equations. The simplest among them, discussed in sections \ref{s2}-\ref{s4}, have close similarity with the laws of thermodynamics and have led to deeper understanding, as well as puzzles, at the interface of general relativity, quantum physics and statistical mechanics. In the dynamical regime, detailed insights on horizon mechanics have come from the infinitely many `balance laws' that relate the time derivatives of multipoles to the fluxes of geometric fields across $\DH$. While these balance laws do not have direct thermodynamical analogs, they provide powerful tools to probe how horizons evolve and to develop gravitational wave tomography. Use of quasi-local horizons is essential in these applications; there are fundamental conceptual difficulties in extending the constructions and results to EHs. Similarly, since the notions of IHs and DHs make no reference to infinity, these frameworks can be used also in spatially compact space-times. The notion of an event horizon, by contrast, does not naturally extend to these space-times. 

The results we summarized have been extended to allow presence of a cosmological constant $\Lambda$. Moreover, in the $\Lambda <0$ case, a difficulty with the first law  pointed out by Caldarelli et al 2000 (in 4 dimensions) and Gibbons, et al 2005 (in higher dimensions) does not arise in the IH framework (Ashtekar et al 2006). For DHs, the only significant change is that the topology of cuts $\S$ of DHs is restricted to be $\S^2$ if $\Lambda >0$ and is completely unrestricted if $\Lambda <0$. For EHs and IHs, results have also been extended to higher dimensions. For DHs, the issue remains open. Similarly, while the first law has been generalized to globally stationary space-times in a wide class of diffeomorphism covariant theories (Iyer \& Wald 1994), such a generalization has not been carried out for IHs in non-stationary space-times. Hamiltonian treatment of the first law with an IH as the inner boundary has been available for some time in the covariant phase space framework. Since DHs are space-like, one can use standard 3+1 Hamiltonian framework. In fact Eq. (\ref{area-increase}) is precisely a linear combination of constraints. However, a fully satisfactory covariant phase space framework in which $\DH$ appears as an inner boundary is not yet available. Such a framework could be very useful to efforts aimed at extending tomography to the fully non-linear regime, beyond the quasi-normal mode approximation.

Finally, due to space limitation this article focuses only on classical gravity. The quasi-local horizon framework also plays a key role in the analysis of the Hawking evaporation process beyond the external field approximation. There are strong indications from the analysis of the tractable  models that what forms in the gravitational collapse and evaporates quantum mechanically are quasi-local horizons. During the collapse, a dynamical horizon forms, which then turns into a time-like world tube of shrinking marginally trapped surfaces during quantum evaporation because the energy flux into the horizon is negative. In the semi-classical regime while we have these quasi-local horizons, one cannot even ask if there is an event horizon because $\scrip$ is incomplete. For a summary, including expectations beyond the semi-classical regime, see, e.g. Ashtekar 2020.


\section*{Acknowledgments}

I have benefitted from discussions with a large number of colleagues. I would especially like to thank Badri Krishnan, Jerzy Lewandowski, Ivan Booth, Steve Fairhurst, Jos\'e Luis Jaramillo and Neev Khera.

\section*{References}

\ni DeWitt B S and DeWitt C M (eds) (1972) \emph{Black holes},
(North-Holland, Amsterdam)

\ni Hawking S W and Ellis G F R (1973) \textit{Large scale
structure of space-time} (Cambridge UP, Cambridge)

\ni Bardeen J W, Carter B and Hawking S W (1973) The four laws of
black hole mechanics, \textit{Commun.\ Math. \ Phys.} \textbf{31}
161

\ni Geroch R and Horowitz G T (1978) Asymptotically simple does not imply asymptotically Minkowskian \textit{Phys. Rev. Lett.} \textbf{40}, 203

\ni Iyer V and Wald R M (1994) Some Properties of Noether charge and a proposal for dynamical black hole entropy. \textit{Phys. Rev.} \textbf{D50} 846-64; \texttt{arXiv:gr-qc/9403028}

\ni Wald R M (1994)  \textit{Quantum field theory in curved spacetime and black hole thermodynamics} (University of Chicago Press) 

\ni Frolov V P and Novikov I D (1998) \textit{Black hole physics}(Kluwer, Dordrecht)

\ni Ashtekar A, Fairhurst S and Krishnan, B (2000) Isolated Horizons: Hamiltonian Evolution and the First Law, \textit{Phys. Rev.} \textbf{D62} 104025; \texttt{gr-qc/0005083}.

\ni Caldarelli M M, Cognola G, and Klemm D (2000) Thermodynamics of Kerr-Newman-AdS Black
Holes and Conformal Field Theories, \textit{Class. Quant. Grav.} \textbf{17} 399-420; \texttt{arXiv:hep-th/9908022}

\ni Ashtekar A, Beetle C and Lewandowski J (2001) Mechanics of rotating black holes, \textit{Phys. Rev.} \textbf{64} 044016; \texttt{gr-qc/0103026}

\ni Ashtekar A, Beetle C and Lewandowski J (2002) Geometry of generic isolated horizons \textit{Class. Quant. Grav.}\textbf{19} 1195--1225; \texttt{gr-qc/0111067}

\ni Ashtekar A and Krishnan B (2003) Dynamical horizons and their properties, \textit{Phys. Rev.} \textbf{D68} 104030; \texttt{gr-qc/0308033}

\ni Ashtekar A and Krishnan B (2004) Isolated and dynamical horizons and their applications, \textit{Living Rev. Rel.} \textbf{10} 1--78; \texttt{gr-qc/0407042}

\ni Ashtekar et al (2004) Multipole Moments of Isolated Horizons, \textit{Class. Quant. Grav.}, \textbf{21} 2549-2570; \texttt{gr-qc/0401114}

\ni Gibbons G W, Perry M J and Pope C N (2005) The first law of thermodynamics for Kerr-anti-de
Sitter black holes \textit{Class. Quant. Grav.} \textbf{22} 1503-1526; \texttt{arXiv:hep-th/0408217}

\ni Ashtekar A and Galloway G (2005) Some uniqueness results for dynamical horizons, \textit{Adv. Math. Theo. Phys.} \textbf{9}, 1--34; \texttt{gr-qc/0503019}

\ni Ashtekar A, Pawlowski T, Van Droeck C (2006) Mechanics of higher-dimensional black holes in asymptotically anti-de Sitter space-times \textit{Class. Quant. Grav.} \textbf{24} 625--644; \texttt{arXiv:gr-qc/0611049} 

\ni Andersson L, et al 
(2009) The time evolution of marginally tapped surfaces \textit{Class. Quant. Grav.} \textbf{26} 085018; \texttt{arXiv:0811.4721}

\ni Jaramillo J. L. (2012) An introduction to local black hole horizons in the 3+1 approach to general relativity, \textit{Int.J. Mod. Phys.} Conf. Series \textbf{7}, 31-66, \texttt{arXiv:1108.2408}

\ni Ashtekar A, Campiglia, M and Shah, S. (2013) Dynamical black holes: Approach to the final state, \textit{Phys. Rev.} \textbf{D88} 064045; \texttt{arXiv:1306:5697}


\ni Gupta, A, et al (2018) Dynamics of marginally trapped surfaces in a binary black hole merger: Growth and approach to equilibrium, \textit{Phys. Rev.} \textbf{D97}, 084028; \texttt{1801.07048}

\ni Pook-Kolb D, et al (2019) Interior of a Binary Black Hole Merger, \textit{Phys. Rev. Lett.} \textbf{123}, 171102; \texttt{arXiv:1903.05626}

\ni Prasad V, et al (2019) News from horizons in binary black hole mergers, \textit{Phys. Rev. Lett.} \textbf{125} 121101; \texttt{arXiv:2003.06215}

\ni Giesler M, et al (2019) Black hole ringdown: The importance of overtones \textit{Phys. Rev.} \textbf{X9} 041060; \texttt{arXiv:1903.08284}

\ni Ashtekar A (2020) Black Hole evaporation: A perspective from loop quantum gravity, \texttt{Universe} \textbf{6} 21; \texttt{arXiv:2001.08833}
  
\ni Pook-Kolb D, et al (2020) Horizons in a binary black hole merger II: Fluxes, multipole moments and stability; \texttt{arXiv:2006.03940} 

\ni Pook-Kolb D, Booth I, and Hennigar R A (2021) What Happens to Apparent Horizons in a Binary Black Hole Merger?  \textit{Phys. Rev. Lett.} \textbf{127} 181101; \texttt{arXiv:2104.10265}

\ni Ashtekar A, et al (2022a) Non-expanding horizons: Multipoles and symmetries, \textit{JHEP} \textbf{01}, 28; \texttt{arXiv:2111.07873}

\ni Ashtekar A, et al (2022b) Charges and fluxes on (perturbed) non-expanding horizons, \textit{JHEP} \textbf{02} 066; \texttt{arXiv:2112.05608}

\ni Prasad V, et al (2022) Tidal deformation of dynamical horizons in binary black hole mergers, \textit{Phys. Rev.} \textbf{D105} 044019; \texttt{arXiv:2106.02595}

\ni Chen Y,  et al (2022) Multipole moments on the common horizon in a binary-black-hole simulation, \textit{Phys. Rev.} \textbf{D106}  124045; \texttt{arXiv:2208.02965}

\ni Gadioux G and Reall H (2023) Creases, corners and caustics: properties of non-smooth structures on black hole horizons, \texttt{arXiv:2303.15512}

\ni Baibhav V, et al (2023)  Agnostic black hole spectroscopy: quasinormal mode content of numerical relativity waveforms and limits of validity of linear perturbation theory; \texttt{arXiv:2003.06215}

\ni Khera N, et al (2023) Nonlinearities at black hole horizons, \texttt{arXiv:arXiv:2306.11142}

\end{document}